\documentclass[11pt]{article}

\usepackage{graphics,epsfig}
\usepackage{amsmath}
\usepackage{amsfonts}
\usepackage{amssymb}

\newcommand{\no}{\nonumber\\}
\newcommand{\be}{\begin{equation}}
\newcommand{\ee}{\end{equation}}
\newcommand{\ba}{\begin{eqnarray}}
\newcommand{\ea}{\end{eqnarray}}

\newcommand{\la}[1]{\label{#1}}

\date{}

\begin{document}
\title{Dilepton excess  from local parity breaking in baryon matter}
\author{A. A. Andrianov$^{a,b}$
\footnote{Corresponding author: E-mail: andrianov@icc.ub.edu, FAX: +34934037063},
\ V. A. Andrianov$^{b}$,
\ D. Espriu$^{a}$,
\ X. Planells$^{a}$
\\
\small{$^a$\ \ Departament d'Estructura i Constituents
de la Mat\`eria}\\ \small{and Institut de Ci\`encies del Cosmos (ICCUB),}\\
\small{
Universitat de Barcelona, Mart\'\i \ i Franqu\`es 1, 08028 Barcelona, Spain}\\
\small{$^b$\ \  V. A. Fock Department of Theoretical Physics,}\\
\small{Saint-Petersburg State University, 198504, St. Petersburg, Russia}}
\maketitle

\begin{abstract} We propose that local parity breaking induced by a large-scale fluctuation of
topological charge at large temperatures
and/or condensation of pseudoscalar mesons in the isotriplet channel for large baryon densities may be
responsible for the substantial dilepton excess that is found for low invariant masses and moderate values of $p_T$
in central heavy ions collisions.
This insofar unexplained enhancement could be understood by a combination
of two effects leading both to an excess of $e^+e^-$ and $\mu^+ \mu^-$  pairs with respect to
theoretical predictions based on conventional hadronic processes: (a) a  modification of the
dispersion relation of photons and vector mesons propagating in such a nuclear medium due to local parity breaking;
(b) the appearance of new decay channels, forbidden by parity conservation in QCD in the usual vacuum.
Possible signatures of this effect and perspectives for its detection are discussed.
\end{abstract}

\section{Introduction}
The possibility of local parity breaking (LPB) in baryon matter at extreme conditions has been
discussed recently \cite{kharzeev,zhit,anesp,avep} in association with relativistic heavy ion collisions (HIC) at RHIC,
GSI, and CERN \cite{reviews}.
It has been suggested in \cite{kharzeev}
that at high temperatures an isosinglet pseudoscalar background could
arise due to large-scale topological charge fluctuations (studied recently in
lattice QCD simulations \cite{lattice}). These considerations led eventually to the
formulation of the so-called Chiral Magnetic Effect (CME)\cite{kharzeev}, claimed
to have already been
detected in the STAR and PHENIX experiments at RHIC \cite{star}. The effect should be most visible
for noncentral HIC where large angular momenta induce large magnetic fields contributing
to the chiral charge separation. In a separate development it was proposed in \cite{anesp} that
the presence of a phase where parity was spontaneously broken could be a rather generic feature of QCD at
finite density. This conclusion was reached using effective Lagrangians; a proof of principle is
unfortunately still missing in QCD due to the difficulties of dealing with non-vanishing chemical
potentials in lattice QCD.

On the other hand, in central HIC an abnormal
dilepton excess has been observed \cite{NA60,phenix} (past experiments are reviewed in \cite{reviews}).
In the PHENIX experiment, for instance, an excess is seen for
dileptons in the range $ M  < 1.2$ GeV, for centrality $0 \div 20\%$ and $p_T < 1$ GeV \cite{phenix}.
Theorists have so far been unable to account for this dilepton excess convincingly. Thermal effects inducing vector
resonance broadening and/or  mass dropping fall short of providing a full explanation \cite{rapp}-\cite{zahed}.

The dilepton excess is conspicuously absent for peripheral HIC (where the CME should be more visible).
We conjecture that the CME and the `anomalous' dilepton excess may be complementary effects revealing
two facets associated to the formation of a thermodynamic phase where parity is locally broken. This is the
main contention of this work.

We shall assume the emergence a of time-dependent pseudoscalar background
(associated with axial chemical potential induced by a non-trivial topological charge \cite{kharzeev,zhit}
or with pseudoscalar condensation \cite{anesp}) in central HIC\footnote{We could refer
the reader to \cite{axion} where it has been shown
that a pseudoscalar field slowly evolving in time drastically changes the
electromagnetic properties of the vacuum. In particular a photon propagating in this background
with sufficiently high energy may decay on shell in medium
into dileptons.} and search for possible manifestation of local parity breaking (LPB) in electromagnetic
probes. We investigate whether these changes may induce a large dilepton excess in certain HIC.
Without excluding the relevance of more conventional mechanisms to partially account for the enhancement,
we shall conclude that a good part of the excess of dileptons might be a consequence of LPB due to the generation
of a pseudoscalar, isosinglet or neutral isotriplet, condensate whose precise magnitude and time variation depends on
the dynamics of the HIC.

This paper is mostly concerned with the possibility of LPB associated to the appearance of local fluctuations
of the topological charge (or equivalently, as we shall argue below, of a axial chemical potential). The issue of
isotriplet condensation induced by finite density\footnote{Finite density may contribute to isosinglet
pseudoscalar condensation too.} will
not be discussed here (see \cite{anesp} for some tentative considerations).
As we will see, this form of LPB can be incorporated by adding a parity-odd term in the Lagrangian; however, this
breaking, while being technically `soft', can be numerically large, leading to important consequences.
We shall review in detail how in-medium hadronic resonances are influenced by this effect and
discuss possible ways to verify its existence. We will briefly discuss how hadronic processes are substantially modified
by the presence of the parity breaking piece and point out their likely relevance for the dilepton enhancement.
However a detailed discussion of the latter aspects as well as a complete comparison with the
experimental data are postponed to a subsequent publication.

\section{Generation of axial baryon charge and axial chemical potential}
Let us outline the relationship between emergent topological charge, baryon axial charge and axial chemical potential.
Let us assume that a jump of topological charge $T_5$ localized in a finite
volume  arises (for instance, in a hot environment due to
sphaleron transitions \cite{sphaleron,sphaleronqcd}) as a consequence of
a HIC, and survives for a sizeable lifetime in the fireball,
$\Delta t \simeq \tau_{\mbox{fireball}}\simeq 5-10\ $ fm.
For a given gauge field $G_\mu(x)$ the topological charge itself
\be T_5(t)=\frac{1}{4\pi^2}\int_{\text{vol.}}d^3x \, K_0,\quad K_\mu =  \frac12 \epsilon_{\mu\nu\rho\sigma}\text{Tr}\left(G^\nu\partial^\rho G^\sigma -i\frac23 G^\nu G^\rho G^\sigma\right), \ee
is not gauge invariant under large gauge transformations. Conventional notations for the gluon field and the
gluon field-strength are used \cite{kharzeev}. However, the jump in $T_5$ can be associated to the
space-time integral of the gauge-invariant Chern-Pontryagin density
\ba
\Delta T_5 = T_5(t_f) - T_5(0) = \frac{1}{16\pi^2}\int^{t_f}_0 dt\int_{\text{vol.}}d^3x \, \text{Tr}
\left (G^{\mu\nu} \widetilde G_{\mu\nu}\right ) = \frac{1}{4\pi^2}\int^{t_f}_0 dt\int_{\text{vol.}}d^3x \, \partial^\mu K_\mu .
\ea
We shall neglect here the topological current flux through the fireball boundary during the thermodynamic phase.

It is well known that for the color- and iso-singlet axial current $J_{5,\mu} =\bar q \gamma_\mu\gamma_5  q$
the local PCAC relation is afflicted with a gluon anomaly
\be
\partial^\mu J_{5,\mu} -2 i \bar q \hat m_q \gamma_5 q = \frac{N_f}{2\pi^2} \partial^\mu K_\mu. \label{pcac}
\ee
This exact equation allows to associate a non-zero topological charge with a non-trivial axial charge $Q_5^q$.
Indeed, one can integrate the local PCAC \eqref{pcac} over a finite space volume of fireball where a
nontrivial topological charge is located and find the connection between their time derivatives
\be
\frac{d}{dt}  (Q_5^q - 2 N_f T_5)  \simeq  2 i \int_{\text{\small vol.}} d^3 x\, \bar q \hat m_q \gamma_5 q\ ,
\quad  Q_5^q =\int_{\text{\small vol.}}d^3x\, \bar q\gamma_0\gamma_5q \label{axcons}.
\ee
In the previous equation
the fluxes across the fireball surface of the  axial and topological currents in the absence of strong magnetic fields
are neglected (this would most likely be the case in central HIC where there is no CME). We now assume that the
manifest breaking of axial current conservation, generated by quark masses, can be neglected for small masses and finite volumes
(as there are no proper zero modes then). Then the induced axial charge turns out to be
conserved provided that the topological charge is not changed during $\tau_{\mbox{fireball}}$.

In HIC
one can conceive of the following evolution: at the initial stage the nuclear matter is  compressed and heated
(during a time $\tau_{\mbox{heating}} < 0.5 $ fm) and a topological charge emerges due to a
sphaleron transition\cite{sphaleronqcd}. For light quarks
the  creation of topological charge leads to the generation of an axial charge. At
 the next stage the nuclear matter cools down (for a time $\tau_{\mbox{fireball}} \sim 5-10 $ fm ) until freeze-out.
During this period
the topological charge is supposedly conserved.
The axial charge would be conserved too provided that the quark mass term breaking chiral symmetry remained subdominant.
The characteristic oscillation time is governed by inverse quark masses. Evidently for $u,d$ quarks
$1/\hat m_q \sim 1/5$ MeV$^{-1}$ $\sim 40$ fm $\gg \tau_{\mbox{fireball}}$ and the left-right quark mixing can be neglected.
But it is not the case for strange quarks as  $1/ m_s \sim 1/200$ MeV$^{-1}$ $\sim 1$ fm $\ll \tau_{\mbox{fireball}}$ and even if
a topological charge persists during fireball lifetime, the mean value of strange quark axial charge is around zero
due to  left-right oscillations.

Thus for $u,d$ quarks, QCD with a  topological charge $\langle \Delta T_5 \rangle\neq 0$ can be equally
described at the Lagrangian level by
topological chemical potential $\mu_\theta$ or by axial chemical potential $\mu_5$
\be
\langle \Delta T_5 \rangle \simeq \frac{1}{2N_f} \langle Q_5^q \rangle \, \Longleftrightarrow \, \mu_5 \simeq \frac{1}{2N_f} \mu_\theta,
\ee
simply adding to the QCD Lagrangian $\Delta {\cal L}_{\mbox{\rm top}}= \mu_\theta\Delta T_5$ or, alternatively, $\Delta {\cal L}_q = \mu_5 Q_5^q$.

In a Lorentz invariant form one can associate a  classical background $a(x)$, depending smoothly on time, so that
\be
 \Delta {\cal L}_G = \frac{N_f}{2\pi^2} K_\mu \partial^\mu a(x)  = \frac{1}{4\pi^2}\mu_\theta  K_0\, \Longleftrightarrow\,
\mu_5 \bar q \gamma_0\gamma_5  q;\quad \mu_5 \simeq \dot a(t)\simeq {\rm constant}.
\ee
Thus, we  assume that a time dependent but approximately spatially homogeneous background
of a pseudoscalar field $a(t)$ is induced at the energy
densities reached in HIC during the fireball evolution and its gradient defines a 4-vector
$\zeta_\mu  \simeq \partial_\mu a \simeq (\zeta,0,0,0)$.
The field $a(t)$   could be either $SU(3)_f$ singlet or diagonal part of $SU(3)_f$ octet
or even a mixture of the two.

If we want to investigate LPB in HIC with the help of electromagnetic probes, we have to account
for the photon contribution to the singlet axial anomaly
\be
Q_{5}^q \, \rightarrow \, \tilde Q_5=Q_{5}^q- T^{\text{em}}_5,\quad T^{\text{em}}_5 =\frac{N_c}{8\pi^2}\int_{\text{vol.}}d^3x \, \varepsilon_{jkl} \, \text{Tr}\left (\hat A^j\partial^k \hat A^l\right).
\ee
Now $\mu_5$ is conjugated to the (nearly) conserved $\tilde Q_5$ but not to $Q_5$ itself \cite{rubakov}.

After bosonization of QCD in the light meson sector we use the quark-meson description of $\tilde Q_{5}^q$
with vector meson and photon fields $V_\mu$ appearing in the quark covariant derivative.
The anomalous PCAC relation \eqref{pcac} is bosonized following the Vector Dominance Model (VDM) prescription \cite{vmd,vmd2}.
The extra term in the Lagrangian takes the form \cite{truhlik} $
\Delta\mathcal L\simeq \varepsilon^{\mu\nu\rho\sigma}\text{Tr}\left [\hat\zeta_\mu V_\nu V_{\rho\sigma}\right ],$
with $\hat\zeta_\mu=\hat \zeta\delta_{\mu 0}$ for spatially homogeneous and isotropic fireball
(the hat denotes the isospin content
in matrix notation).
Therefore a non-trivial topological chemical potential $\mu_\theta$  is transmitted via axial chemical
potential to a non-trivial axial charge inducing the photon and meson Chern-Simons Lagrangian
(a similar motivation can be seen in \cite{zhit}). And vice-versa, after measuring the LPB background
$\zeta$ one could, in principle, find the value of the mean topological charge in fireballs.

When nuclear matter is superdense, pseudoscalar condensation in the $I=1$ channel
may occur too, as it has been outlined in \cite{anesp} (and references therein). This situation corresponds to
introducing a vector chemical potential $\mu$. As indicated in the introduction, the appearance of LPB is inferred
in this case from effective Lagrangian techniques, and a direct lattice confirmation is still missing. In spite of this,
the conclusions of the analysis in \cite{anesp}
seem rather robust: there is a range of densities where a LPB vacuum is energetically favorable. Thus
 for light quarks in hot and dense nuclear matter the matrix structure of the Chern-Simons vector $\zeta$ in
flavor space generically includes not only $SU(3)_f$ singlet but also neutral components of $SU(3)_f$ octet, that corresponds to
including both a $\mu_5$ and a $\mu$ term, respectively. We note that the appearance of a condensate in the
$\tau_3$ direction implies in addition a breaking of the isospin symmetry of the vacuum.

\section{VDM Lagrangian in the presence of P-breaking background}
The appropriate framework to describe
electromagnetic interactions of hadrons at low energies is the VDM
containing the lightest vector mesons $\rho^0$, $\omega$ and $\phi$ in the $SU(3)$ flavor sector.
 Quark-meson interactions are described by
\ba
{\cal L}_{\text{int}} = \bar q \gamma_\mu V^\mu q;\quad  V_\mu \equiv - e A_\mu Q  +
\frac12 g_\omega  \omega_\mu \mathbf{I}_{q} + \frac12 g_\rho \rho_\mu  \lambda_3 + \frac1{\sqrt 2}g_\phi \phi_\mu \mathbf{I}_{s}, \label{veclagr}
\ea
where
$Q= \frac{\lambda_3}{2} + \frac16 \mathbf{I}_{q} - \frac13 \mathbf{I}_{s}$, $g_\omega \simeq  g_\rho \equiv g \simeq 6 < g_\phi \simeq 7.8$;
 $\mathbf{I}_{q}$ and $\mathbf{I}_{s}$ are the identities in the non-strange and strange sector, respectively; and $\lambda_3$ is
the corresponding Gell-Mann matrix. The Maxwell and mass terms are
\ba\label{vdm}
&&\!\!\!{\cal L}_{\text{kin}} = - \frac14 \left(F_{\mu\nu}F^{\mu\nu}+ \omega_{\mu\nu}\omega^{\mu\nu}+
 \rho_{\mu\nu}\rho^{\mu\nu} +
 \phi_{\mu\nu}\phi^{\mu\nu}\right)+ \frac12  V_{\mu,a}  m^2_{ab} V^\mu_b, \no
&&\!\!\! m^2_{ab} =
m_V^2\left(\begin{array}{ccccccc}
\frac{4 e^2}{3g^2} & &-\frac{e}{3g} && -\frac{e}{g}&& \frac{\sqrt 2 eg_\phi}{3g^2}  \\
 -\frac{e}{3g}&& 1 && 0 &&0 \\
 -\frac{e}{g} && 0 && 1 &&0 \\
 \frac{\sqrt 2 eg_\phi}{3g^2}  && 0  && 0  &&\frac{g_\phi^2}{g^2} \\
\end{array}\right),\ \mbox{\rm det}\left( m^2\right) = 0,\nonumber
\ea
where $(V_{\mu,a})
\equiv \left(A_\mu,\, \omega_\mu, \, \rho_\mu^0 \equiv \rho_{\mu}, \, \phi_\mu\right)$ and  $m_V^2 = m^2_\rho = 2 g^2_\rho f_\pi^2\simeq m^2_\omega$ . This matrix reflects the VDM relations
at the quark level \cite{vmd,vmd2}. Finally, in a pseudoscalar time-dependent background
the Lagrangian contains a parity-odd Chern-Simons (CS) term
\ba
{\cal L}_{\text{CS}}(k)\,= - \frac14 \varepsilon^{\,\mu\nu\rho\sigma}\, \mbox{\rm Tr}\left [ \,\hat\zeta_\mu \, V_\nu(x)\, V_{\,\rho\sigma}(x)\right ]= \frac12 \mbox{\rm Tr}\left [\,\hat\zeta \,\epsilon_{jkl}\, V_{j} \,\partial_k V_{l} \right ]
= \frac12 \,\zeta\,\epsilon_{jkl}\, V_{j,a} \,N_{ab}\,\partial_k V_{l,b}, \label{CS}
\ea
which additionally mixes photons and vector mesons due to LPB. With this definition,
the relation $\zeta=N_c\;g^2\mu_5/8\pi^2$ holds.
Numerically $\zeta \simeq 1.5 \mu_5$.

At this point, one should recall the arguments given in the section 2 about the fact that the strange quark flavor
approximately decouples from the rest due to intensive left-right oscillations as $m_s \gg \tau_{\mbox{fireball}}$. In consequence,
even if a LPB condensate appears in the strange sector, its average will be essentially zero.
In addition the role of $\phi$ meson is negligible as its typical mean free path $\sim 40$ fm makes it less sensitive
to medium effects.
In this sense,
from now on we shall consider a general $\hat \zeta$ to be a linear combination of isosinglet and isotriplet cases.
Therefore only the flavor matrices $\mathbf{I}_{q}$ and $\lambda_3$ are relevant and
the $\phi$ meson will not be considered in the mass matrix.

For isosinglet pseudoscalar background $\hat\zeta=\mathbf{I}_{q} 2\zeta/{g^2}$, the mixing matrix reads
\ba
(N_{ab}^\theta) \, \simeq\,  \left(\begin{array}{ccccc}
\frac{10e^2}{9g^2} &  &-\frac{e}{3g}& &-\frac{e}{g}\\
-\frac{e}{3g}& & 1 & & 0\\
-\frac{e}{g}& & 0 & & 1\\
\end{array}
\right)=\frac{(m^2_{ab})}{m_V^2}\Bigg|_{SU(2)_f},\ \ \mbox{\rm det} (N^\theta) = 0 . \la{tab1}
\ea
This case is expected to be the dominant one in experiments where the temperature in the nuclear fireball
is much larger than the chemical potential $T\gg\mu$, as it is the case for the LHC or RHIC\footnote{Experiments at FAIR
and NICA will explore the region $\mu \gg T$ in the future}.

Let us now try to estimate the value of $\zeta$ using simple considerations. A first naive estimate would be to assume
simply that $\zeta\sim \tau^{-1}$ so a formation time of the pseudoscalar condensate
$\tau\sim 0.5$ fm leads to $\zeta\sim 400$ MeV. On the other hand, we can estimate $\mu_5$ by assuming that
the change in the free energy has to be of order $f_\pi$ and the average change in the axial charge $\langle
\Delta Q_5\rangle\sim 1 $. Then, assuming a linear response, $\mu_5 \sim f_\pi$ and $\zeta\sim 150 $ MeV. We conclude
that the natural value for $\zeta$ is in the few hundreds of MeV. That is to say that its effects on low-energy
hadronic physics are potentially large and let us see how its presence affects the in-medium
meson dispersion relations.

The mass-shell equations for vector mesons read
\ba
\!\!\!\! K^{\mu\nu}_{ab} V_{\nu, b} = 0;\quad k^\nu\,V_{\nu, b} = 0, \quad
K^{\mu\nu} \equiv g^{\mu\nu} (k^2 \mathbf{I} -  m^2) - k^\mu k^\nu \mathbf{I} -
i \varepsilon^{\,\mu\nu\rho\sigma}\,\zeta_\rho  k_\sigma  N^\theta ,
\ea
and select three physical polarizations vectors $(\varepsilon_+,\varepsilon_-,
\varepsilon_L)$ for massive vector fields that couple to conserved fermion currents (the polarization projectors are described in \cite{axion}).
The longitudinal polarization $\varepsilon^\mu_L$ is orthogonal to $k_\mu$ and to  $\zeta_\mu$
The mass of this state remains undistorted while the transversal
polarizations satisfy
\ba
K^{\mu}_{\nu}\varepsilon^\nu_\pm = \Big(k^2 \mathbf{I} -  m^2 \pm \sqrt{(\zeta \cdot k)^2 - \zeta^2 k^2}\ N \Big) \varepsilon^\mu_\pm;\qquad
 m^2_{V,\pm}\equiv k_0^2 - \vec k^2 \simeq m_V^2 \mp \zeta |\vec k| . \label{mvec}
\ea
The spectrum is found after simultaneous diagonalization of $ m^2|_{SU(2)}\sim N^\theta $ with $\zeta_\mu \simeq (\zeta, 0,0,0)= $
constant.
We notice that in the case of pure isosinglet pseudoscalar background massless photons are not distorted
when mixed with massive vector mesons.
In turn massive vector mesons split into three polarizations with masses $m^2_{V,+} < m^2_{V,L}< m^2_{V,-}$.
This splitting unambiguously signifies parity breaking as well as violation of Lorentz invariance
(due to the time-dependent background).
Note that the position of resonance poles for $\pm$ polarized mesons is moving with wave vector $|\vec k|$ and
therefore they appear as broadened resonances, leading
to an enhancement of their contribution to dilepton production away from their nominal
vacuum resonance position.

\section{Parameters of the VDM Lagrangian and new processes}

The VDM coefficients in \eqref{vdm} and \eqref{tab1}
are obtained from the anomalous Wess-Zumino action \cite{truhlik}
and related to the phenomenology of radiative decays of vector mesons \cite{radec}.
In particular the VDM coefficients in \eqref{tab1} can be estimated from the experimental
decay constants \cite{pdg} for the processes
$\eta \rightarrow \gamma\gamma$, $\eta' \rightarrow \gamma\gamma$,
$\omega \rightarrow \eta\gamma$, $\rho^0 \rightarrow \eta\gamma $
after taking into account  a strong $\eta_8 - \eta_0$ mixing \cite{mixing}.
Only the ratio of the decay widths $\omega \rightarrow \eta\gamma$, $\rho^0 \rightarrow \eta\gamma $
is a little sensitive to the mixing and confirms the off-diagonal elements of \eqref{tab1}.
Armed with the above information and after estimating $\zeta$ one can easily determine the modifications
on the vector meson spectrum and their eventual relevance for the dilepton excess.

However, this is not the end of the story. If there is parity breaking
many other processes are possible. They can be estimated by using the spurion technique;
i.e.  $\mu_5$ as the time component of a fictitious axial field. Two
new processes are then likely to be most relevant inside the fireball thermodynamics: the decays
$\eta,\eta^\prime \to \pi\pi$ that are strictly forbidden in QCD on parity grounds.

To get a rough estimate of the relevance of these previously forbidden processes we take the parity even sector
and extend the covariant derivative  by including a axial
chemical potential, $D_\nu \Longrightarrow  D_\nu - i \{\mu_5 \delta_{0\nu}, \cdot \}=D_\nu - 2i\mu_5 \delta_{0\nu}$.
There is no contribution from dimension two operators, but dimension four terms (see e.g. \cite{chpt})
lead to
\be
\sim \frac{16\mu_5}{F_\Pi f_\pi^2} L \, \partial \eta \, \mbox{Tr}\left( \partial\hat\pi \partial\hat\pi\right);
\quad \Pi = \eta,\, \eta',
\ee
where $L$ is a combination of the Gasser-Leutwyler constants $L_{1,2,3} \sim 10^{-3}$.
To get a numerical estimate we take the average pion momenta
to be $\sim m_\eta/2$. The effective coupling constant affecting this operator can be estimated to be $\sim 0.4$,
which is large enough to induce substantial $\eta$ meson regeneration in the hot pion gas.
A very rough estimate of the partial width
for the exotic process under discussion gives
$\Gamma_{\eta\to\pi\pi}\simeq 100\,\text{MeV} $,
when we assume $\zeta\simeq 200$ MeV, to be compared for instance to  $\Gamma_{\rho\to\pi\pi}\simeq 150$ MeV.
Clearly if $\rho$'s are in thermal equilibrium in the pion bath, so will the $\eta$. A similar analysis leads to an
even larger width for the $\eta^\prime$ due to the above parity breaking operator.
On the other hand, the $\omega$ decays are not modified by a
parity breaking isosinglet spurion (but they would for an isotriplet one).

However at this point, one should be aware that the previous estimate using the chiral Lagrangian may not be reliable at all
because the numerical value of the parity breaking terms is quite large.
Then one expects substantial mixing with the scalar partners
of $\eta,\eta^\prime$  (i.e. $\sigma$ and $f_0$) with comparable masses. A detailed analysis
will be given elsewhere. On the contrary
we do not expect substantial mixing between $\rho$ and $a_1$ for an isosinglet condensate.

We have to retain the very important point that a number of light hadronic states will be in
thermal equilibrium (so their respective abundances
will be governed by the Boltzmann distribution) and regeneration
of these resonances will take place in the fireball similarly to the one taking place for $\rho$'s. They
will be much more abundant in the fireball if there is LPB and the decays of these resonances
will be an important source of dileptons so far unaccounted for.

\section{Dilepton production rate in P-breaking medium}

While during the first stage of the HIC, nuclear matter is being compressed and heated
(during $\tau_{\mbox{heating}}$) and a parity breaking condensate $ a(t)$ appears and grows, during the expansion
of the fireball this condensate dilutes during a time $\tau_{\mbox{fireball}}$. One can approximate both regimes
by a linear function with slopes $\sim \zeta_{\mbox{heating}}$ and  $\sim \zeta_{\mbox{fireball}}$, with opposite signs.
Of course, the cooling time greatly exceeds the time
taken for the LPB phase to form so most of the expected effects should come from this latter period.
For the very LPB effect only $|\zeta|$ is
important because a change of sign causes an interchange in polarizations but does not
affect the splitting of masses for different polarizations.
Naturally there is some error in approximating the time dependence of the condensate with a linear parametrization
(which is the one where our results are strictly valid) but qualitative aspects should be well captured assuming
an average or effective constant value for $\zeta$.

Dileptons are produced in a variety of decays. The dominant processes in the range of invariant masses
$200$ MeV $< M < 1200$ MeV are assumed to be the $\rho$ and $\omega$ decays into lepton pairs,
the $\omega\to \pi^0 \ell^+\ell^-$ Dalitz decay and similar
Dalitz decays for $\eta$ and $\eta^\prime$. These are the basic ingredients of the
so called `hadronic cocktail' \cite{rapp}, conventionally used to predict (unsuccessfully for central HIC)
the dilepton yield\footnote{In addition there is a substantial $c\bar c$ background that
is not affected by the present considerations.}. Naturally, the dispersion
relations of $\rho$, $\omega$, etc. are modified due to conventional in-medium effects but
these modifications are insufficient to explain the abnormal dilepton yield.

LPB modifies the calculation in two ways. On the one hand generates new in-medium effects on resonances
that can be unambiguously predicted in term of the parity breaking parameter $\zeta$ alone. In addition,
some hadronic states mix as a consequence
of the parity breaking effect and the new allowed interactions are able to thermalize some of
the lowest lying states, whereas only $\rho$'s could reasonably be expected to be in
thermal equilibrium with the pion gas without LPB. In what follows we will consider the modifications in the dilepton
spectrum due to the $\rho$ and $\omega$
and shall postpone discussing the effects of the new processes allowed by LPB.

The production rate of lepton pairs takes a form similar to the one
given in \cite{rapp} but with modified form factors due to LPB, according to our previous discussion
\begin{align}\label{eleven}
\frac{dN}{d^4xd^4kd^2\vec p_T}=&c_V\frac{\alpha^2}
{48\pi^2 M^2}\left (1-\frac{n_V^2 m_\pi^2}{M^2}\right)^{3/2}\sum_{\epsilon=L,\pm}\frac{1}
{|E_kp_\parallel-k_\parallel E_p|}\\
\nonumber \times& \frac{1}{e^{M_T/T}-1}P_\epsilon^{\mu\nu}\left (M^2g_{\mu\nu}+4p_\mu p_\nu\right ) \dfrac{m_{V,\epsilon}^4\left (1+\frac{\Gamma_V^2}{m_V^2}\right )}{\left (M^2-m_{V,\epsilon}^2\right )^2+ m_{V,\epsilon}^4\frac{\Gamma_V^2}{m_V^2}},
\end{align}
where $n_V=2,0$ for $\rho$ and $\omega$ cases respectively, and $M>n_Vm_\pi$. $M_T$ is the transverse mass $M_T^2=M^2+k_T^2$ while
$\vec k_T$ and  $k_\parallel$ are the perpendicular and parallel components, respectively.
The projectors $P_\epsilon^{\mu\nu}$
are detailed in \cite{axion}. A simple thermal average with the Boltzmann distribution  has been included \cite{maclerran},
$T$ being an effective temperature\cite{phenix}. Finally, the constants $c_V$ normalize the contribution of the respective
resonances.

\begin{figure}[h!]
\centering
\vspace{-1em}
\hspace{-7em}\includegraphics[scale=0.23]{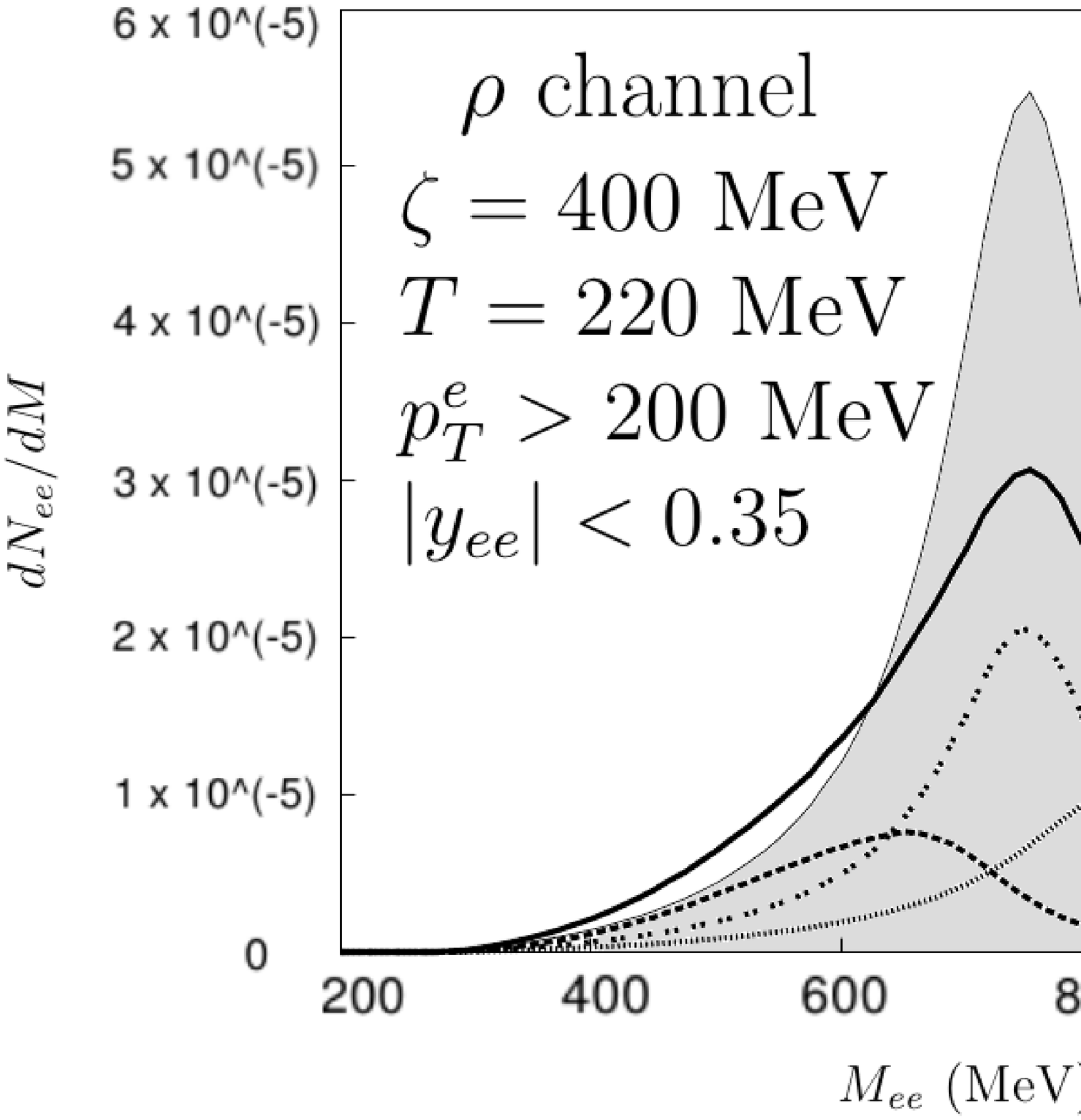}\hspace{10em}\includegraphics[scale=0.23]{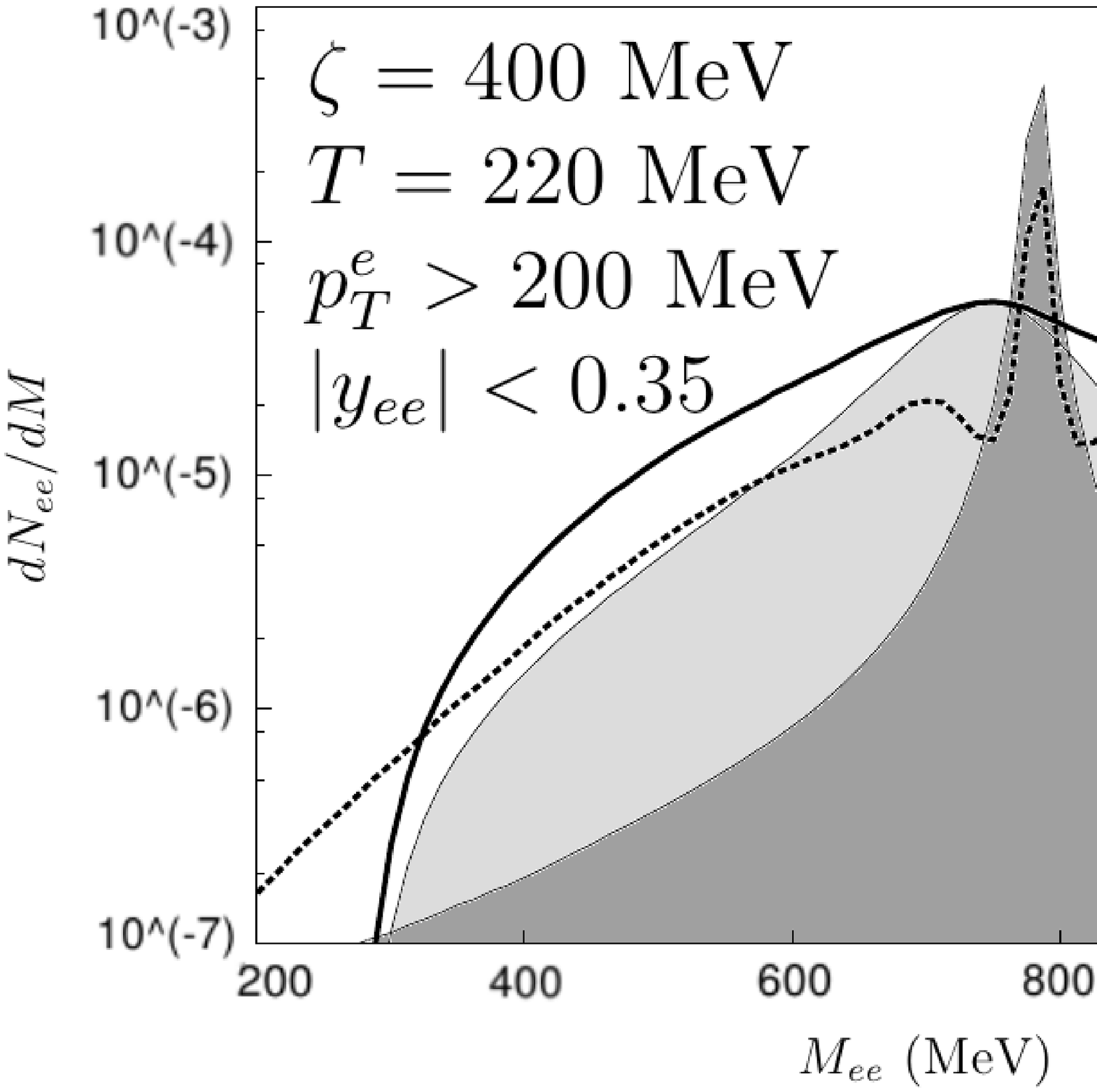}
\vspace{-5em}
\textsc{\caption{\textup{Left: The polarization splitting of the $\rho$ contribution to dilepton production is shown
for LPB with $\zeta = 400$ MeV. The comparison with the case $\zeta = 0$ (shaded region) is presented. Right: the in-medium
contribution in the $\rho$ and $\omega$ channels (solid and dashed line, respectively) is presented for $\zeta=400$ MeV
together with their vacuum contributions (light and dark shaded regions, respectively). The in-medium $\rho$ yield is
enhanced by a factor 1.8 (see text). The vertical units
are taken to coincide with PHENIX experimental data \cite{phenix}, as well as experimental
detector cuts and temperature.}}\label{polasym}}
\vspace{-0.5em}
\end{figure}

Being a theoretical paper, we will not pursue a detailed comparison with experiments. However, in order to make our plots
meaningful we have implemented some values relevant for the PHENIX experimental set-up. For instance since the $\omega$
is not in thermal equilibrium inside the fireball, its relative
weight $c_\omega$ is normalized to the peripheral HIC result (agreeing with existing
$pp$ and $p$-nucleus data \cite{reviews,phenix})
and a relative enhancement of 1.8 is assumed for the $c_\rho/c_\omega$ ratio based on the data (\cite{phenix}; V.L. Rykov, private communication). Likewise
the integration of \eqref{eleven} is performed over the acceptance region of the PHENIX experiment ($p_T>200$ MeV and $|y_{ee}|<$0.35)
and $T\simeq 220$ MeV.

Then one is lead to the dilepton production result that is shown in Figure \ref{polasym} (left), where the $\rho$ spectral
function is presented together with the separate contributions for each polarization for $\zeta=400$ MeV and a
comparison with $\zeta=0$ (no LPB). Similar results may be obtained modifying this parameter, so when $\zeta$ increases,
the circularly polarized resonances appear to be more separated from the vacuum one. Therefore a
measure of dilepton polarization event-by event may reveal in an unambiguous way the existence of LPB,
confirming the hypothesis of pseudoscalar condensate formation in HIC.

As already mentioned for central HIC, particularly at low $p_T$, the $\rho/\omega$ production ratio needs to be enhanced
by a factor 1.8 in PHENIX \cite{phenix}. This enhancement reflects the multiple regeneration
of $\rho$ mesons through $\pi\pi$ fusion in hot pion gas.
There is no such a regeneration for narrow resonances $\omega$ and $\phi$.
The simulation of $\omega$ meson production \cite{phenixomega} shows that a significant fraction
of them decay inside of the nuclear fireball and therefore LPB distorted $\omega$ mesons may also be responsible for
a fraction of the abnormal production of dileptons. This is shown in the right panel of Fig. \ref{polasym}.
In this plot, the separate contributions of $\rho$ and $\omega$ are presented for $\zeta=400$ MeV to be
compared with their vacuum results.  The effect shows a clear
increase in the dilepton production away from the $\rho-\omega$ peak due to the mass shifting of the
circularly polarized resonances in the LPB phase.

Other processes in the region $200$ MeV $< M < 1200$ MeV, especially relevant for dilepton production below the $\rho-\omega$ resonance,
are the $\omega$, $\eta$ and $\eta^\prime$
Dalitz decays, described by the Kroll-Wada
formula \cite{landsberg} that includes the contribution of vector mesons and it remains valid in the
case of LPB provided that we replace the vector meson masses by the values in (\ref{mvec}) according to
the intermediate meson polarization ($L,\pm$).
We have checked that this contribution shows a substantial enhancement
but this and other hadronic processes relevant for dilepton production will be
discussed in a separate publication. The fact that the decaying meson
is distorted by the medium complicates analytically and numerically the calculation due to the lack of Lorentz invariance
generated by a polarization and momentum dependent meson mass.

\section{Conclusions}
We have explored the consequences of assuming LPB via an isosinglet condensate in HIC. We would like to
emphasize the simplicity of the approach presented here. The fits
presented use the values (effective temperatures, normalizations, etc.) quoted by the experiments themselves.
The only free parameter is $\zeta$, which is expected to depend on the characteristics
of the HIC. It should also be said clearly that the presence of LPB does not preclude
other many body or in-medium corrections \cite{rapp,renkr,cassing,zahed}, as long as they do not
represent double counting.

A clear signal of LPB  would verify that dileptons produced for values of the invariant mass above and below the
$\rho-\omega$ pole are predominantly of opposite circular polarizations in event-by-event measurements. This requires
searching for asymmetries among longitudinal
and transverse polarization for different values of $M$ in event-by-event measurements. A more indirect
verification would be a detailed account of the dilepton enhancement in HIC. In this paper we have worked out in detail
the modification of the $\rho$ and $\omega$ dispersion relations, leaving other effects for a forthcoming article.

We have seen that in presence of  an isosinglet time-dependent
pseudoscalar background the vector meson, $\rho$ and $\omega$, propagators are severely distorted.
We have computed this effect and found that it naturally tends to produce an overabundance of  dilepton pairs in the
$\rho-\omega$ resonance region. We have also shown how LPB induces thermalization in the pion gas of other light resonances and
how this could enhance the dilepton production.
Thus LPB seems relevant to explaining  the PHENIX/CERES/NA60 `anomaly'. More work is needed before definite conclusions can be
drawn but we believe that sufficient evidence is accumulated at present to bring these tentative conclusions to the attention of
the interested readers.
\\ \hspace*{3ex}
We acknowledge the financial support from projects FPA2010-20807,\ 2009SGR502,\ CPAN (Consolider CSD2007-00042).
A. \& V. Andrianov are supported also by Grant RFBR 10-02-00881-a and by SPbSU grant 11.0.64.2010.

\end{document}